\documentstyle[12pt]{article}
\font\tenrm=cmr10
\font\sma=cmr10 at 10truept

\def\al{\alpha}
\def\be{\beta}
\def\ga{\gamma}
\def\de{\delta}
\def\ep{\epsilon}
\def\ve{\varepsilon}
\def\ze{\zeta}
\def\et{\eta}
\def\th{\theta}
\def\vt{\vartheta}
\def\io{\iota}
\def\ka{\kappa}
\def\la{\lambda}
\def\vpi{\varpi}
\def\rh{\rho}
\def\vr{\varrho}
\def\si{\sigma}
\def\vs{\varsigma}
\def\ta{\tau}
\def\up{\upsilon}
\def\ph{\phi}
\def\vp{\varphi}
\def\ch{\chi}
\def\ps{\psi}
\def\om{\omega}
\def\Ga{\Gamma}
\def\De{\Delta}
\def\Th{\Theta}
\def\La{\Lambda}
\def\Si{\Sigma}
\def\Up{\Upsilon}
\def\Ph{\Phi}
\def\Ps{\Psi}
\def\Om{\Omega}
\def\mn{{\mu\nu}}
\def\cl{{\cal L}}
\def\fr#1#2{{{#1} \over {#2}}}
\def\prt{\partial}
\def\ap{\al^\prime}
\def\apt{\al^{\prime 2}}
\def\apth{\al^{\prime 3}}
\def\pt#1{\phantom{#1}}
\def\vev#1{\langle {#1}\rangle}
\def\bra#1{\langle{#1}|}
\def\ket#1{|{#1}\rangle}
\def\bracket#1#2{\langle{#1}|{#2}\rangle}
\def\expect#1{\langle{#1}\rangle}
\def\sbra#1#2{\,{}_{{}_{#1}}\langle{#2}|}
\def\sket#1#2{|{#1}\rangle_{{}_{#2}}\,}
\def\sbracket#1#2#3#4{\,{}_{{}_{#1}}
 \langle{#2}|{#3}\rangle_{{}_{#4}}\,}
\def\sexpect#1#2#3{\,{}_{{}_{#1}}\langle{#2}\rangle_{{}_{#3}}\,}
\def\half{{\textstyle{1\over 2}}}
\def\frac#1#2{{\textstyle{{#1}\over {#2}}}}
\def\ni{\noindent}
\def\lsim{\mathrel{\rlap{\lower4pt\hbox{\hskip1pt$\sim$}}
    \raise1pt\hbox{$<$}}}
\def\gsim{\mathrel{\rlap{\lower4pt\hbox{\hskip1pt$\sim$}}
    \raise1pt\hbox{$>$}}}
\def\sqr#1#2{{\vcenter{\vbox{\hrule height.#2pt
         \hbox{\vrule width.#2pt height#1pt \kern#1pt
         \vrule width.#2pt}
         \hrule height.#2pt}}}}
\def\square{\mathchoice\sqr66\sqr66\sqr{2.1}3\sqr{1.5}3}

\newcommand{\beq}{\begin{equation}}
\newcommand{\eeq}{\end{equation}}
\newcommand{\bea}{\begin{eqnarray}}
\newcommand{\eea}{\end{eqnarray}}
\newcommand{\rf}[1]{(\ref{#1})}

\renewenvironment{thebibliography}[1]
 { \rm
   \begin{list}{\arabic{enumi}.}
    {\usecounter{enumi} \setlength{\parsep}{0pt}
     \setlength{\itemsep}{3pt} \settowidth{\labelwidth}{#1.}
     \sloppy
    }}{\end{list}}

\begin{document}
\titlepage

\begin{flushright}
{IUHET 286\\}
{COLBY 94-08\\}
\end{flushright}
\vglue 1cm

\begin{center}
{{\bf SUPERREVIVALS OF RYDBERG WAVE PACKETS\footnote[1]{\tenrm
Invited talk presented by R.B. at the
Fourth Drexel Symposium on Quantum Nonintegrability,
Philadelphia, PA, September 1994
}
\\}
\vglue 1.0cm
{Robert Bluhm$^a$ and V. Alan Kosteleck\'y$^b$\\}
\bigskip
{\it $^a$Physics Department, Colby College\\}
\medskip
{\it Waterville, ME 04901, U.S.A.\\}
\vglue 0.3cm
{\it $^b$Physics Department, Indiana University\\}
\medskip
{\it Bloomington, IN 47405, U.S.A.\\}
\vglue 0.3cm
\bigskip

\vglue 0.8cm
}
\vglue 0.3cm

\end{center}

{\rightskip=3pc\leftskip=3pc\noindent
The revival structure and evolution of
Rydberg wave packets are studied on a time scale much greater
than the revival time $t_{\rm rev}$.
We find a new level of revival structure and
periodic motion different from that
of the known fractional revivals.
The new sequence of revivals culminates
with the formation of a wave packet that more closely
resembles the initial packet than does the full
revival at time $t_{\rm rev}$.
We refer to such a revival as a superrevival.
We also show that an initial radial wave
packet may be described as a type of squeezed
state known as a radial squeezed state.
Our results apply not only for
hydrogenic wave packets,
but for wave packets in alkali-metal atoms as well in
the context of quantum defect theory.

}

\vskip 1truein
\centerline{\it To appear in the}
\centerline{\it
Proceedings of the Fourth Drexel Symposium
on Quantum Nonintegrability,}
\centerline{\it
World Scientific, Singapore, 1995}

\vfill
\newpage

\baselineskip=12pt

\begin{center}
{{\bf SUPERREVIVALS OF RYDBERG WAVE PACKETS \\}
\vglue 1.0cm
{Robert Bluhm$^a$\footnote[1]{\tenrm Speaker} and
V. Alan Kosteleck\'y$^b$
\\}
\bigskip
{\it $^a$Physics Department, Colby College\\}
\medskip
{\it Waterville, ME 04901, U.S.A.\\}
\vglue 0.3cm
{\it $^b$Physics Department, Indiana University\\}
\medskip
{\it Bloomington, IN 47405, U.S.A.\\}

\vglue 0.8cm
}
\vglue 0.3cm

\end{center}

{\rightskip=3pc\leftskip=3pc\noindent\sma
The revival structure and evolution of
Rydberg wave packets are studied on a time scale much greater
than the revival time $t_{\rm rev}$.
We find a new level of revival structure and
periodic motion different from that
of the known fractional revivals.
The new sequence of revivals culminates
with the formation of a wave packet that more closely
resembles the initial packet than does the full
revival at time $t_{\rm rev}$.
We refer to such a revival as a superrevival.
We also show that an initial radial wave
packet may be described as a type of squeezed
state known as a radial squeezed state.
Our results apply not only for
hydrogenic wave packets,
but for wave packets in alkali-metal atoms as well in
the context of quantum-defect theory.

}

\baselineskip=16pt

\vglue 0.6cm
{\bf\noindent 1. Introduction}
\vglue 0.2cm

If a Rydberg atom is excited by a short-pulsed laser field,
a superposition of states with a spread of energy levels results
\cite{pasaalrz}.
Experiments on such systems
have detected electronic motion
with a periodicity $T_{\rm cl}$ equal to the classical
period of a particle in a keplerian orbit.
However,
the motion is not entirely classical,
as the wave packet disperses with time.
After many Kepler orbits
the wave packet recombines into nearly its original shape at
the revival time $t_{\rm rev}$.
Moreover,
prior to this full revival,
the wave function evolves through a sequence of
fractional revivals,
which consist of distinct subsidiary waves
moving with a period that is a fraction of $T_{\rm cl}$
\cite{pasaalrz,avpnaa,note}.
These fractional revivals have been observed in
time-delayed photoionization
and phase modulation experiments.

In the first part of this talk,
we examine the revival structure and evolution
of hydrogenic Rydberg wave packets
for times much greater than the revival time
\cite{srsr2}.
We then consider the case of radial wave packets and show
that the motion of these wave packets
has features characteristic of squeezed states,
and we outline an approach for a squeezed-state description
\cite{bkrssbkradial}.
In the final section,
we describe how a quantum-defect theory based on
supersymmetry may be used to model
wave packets in alkali-metal atoms.
In this context,
we show that the
dependence on the quantum defects
of the long-term revival times for
Rydberg wave packets in alkali-metal atoms
is different from that of the laser detunings
\cite{compare}.

\vglue 0.4cm
{\bf\noindent 2. Superrevivals of Rydberg Wave Packets}
\vglue 0.2cm

The time-dependent wave function for a hydrogenic
wave packet may be expanded in terms of energy eigenstates as
\beq
\Psi ({\vec r},t) = \sum_{n=1}^{\infty} c_n
\varphi_n ({\vec r}) \exp \left[ -i E_n t \right]
\quad .
\label{wave2}
\eeq
Here,
$E_n = -1/2 n^2$ is the energy in atomic units,
and $\varphi_n ({\vec r})$
represents a generic form of the wave function.
For a circular wave packet,
$\varphi_n ({\vec r}) = \psi_{n,n-1,n-1}({\vec r})$,
whereas for a radial wave packet,
$\varphi_n ({\vec r}) = \psi_{n,1,0}({\vec r})$,
where $\psi_{nlm}({\vec r})$ is a
hydrogen eigenstate of
energy and angular momentum.

Both these types of wave packet are
excited by a short laser pulse.
The laser can be tuned to excite coherently
a superposition of states centered on a mean value $\bar n$
of the principal quantum number.
In what follows we assume that the distribution is strongly
centered around $\bar n$.
We may therefore approximate the square of the weighting
coefficients $c_n$ as a gaussian function
of width $\si$.

If we expand the energy in a Taylor series around the centrally
excited value $\bar n$,
we find that the
derivative terms define distinct time scales that depend
on $\bar n$:
$T_{\rm cl} = {2 \pi}/{E_{\bar n}^\prime} = 2 \pi {\bar n}^3$,
$t_{\rm rev} = {- 2 \pi}/{\fr 1 2 E_{\bar n}^{\prime\prime}}
= \fr {2 {\bar n}} 3 T_{\rm cl}$,
and
$t_{\rm sr} = {2 \pi}/{\fr 1 6 E_{\bar n}^{\prime\prime\prime}}
= \fr {3 {\bar n}} 4 t_{\rm rev}$.
The first time scale,
$T_{\rm cl}$,
is the {\it classical keplerian period}.
It controls the initial behavior of the packet.
The second time scale,
$t_{\rm rev}$,
is the {\it revival time}.
It governs the appearance of fractional and full revivals.
The third time scale,
$t_{\rm sr} \gg t_{\rm rev}$,
is a larger time scale we refer to as the
{\it superrevival time}.
This time scale determines the behavior of the packet
for times greater than $t_{\rm rev}$.

Keeping terms through third order,
and defining the integer index $k=n-{\bar n}$,
we may write the wave function as
\beq
\Psi ({\vec r},t) = \sum_{k=-\infty}^{\infty} c_k
\varphi_k ({\vec r}) \exp \left[ -2 \pi i
\left( \fr {kt} {T_{\rm cl}} -  \fr {k^2 t} {t_{\rm rev}}
+ \fr {k^3 t} {t_{\rm sr}} \right) \right]
\quad .
\label{psi3rd}
\eeq

We have found that at certain times $t_{\rm frac}$
it is possible to expand
the wave function $\Psi ({\vec r},t)$
of Eq.\ \rf{psi3rd} as a series of subsidiary wave functions.
The idea is to express
$\Psi ({\vec r},t)$ as a sum of
wave functions $\ps_{\rm cl}$
with matching periodicities
and with shape similar to that of
the initial wave function $\Psi ({\vec r},0)$.
We find that at certain times
$t_{\rm frac} \approx \fr 1 q t_{\rm sr}$,
where $q$ must be an integer multiple of 3,
the wave packet can be written as a sum of
macroscopically distinct wave packets.
Furthermore,
at these times $t_{\rm frac}$,
we also find that the motion of the wave packet is
periodic with a period
$T_{\rm frac} \approx \fr 3 q t_{\rm rev}$.
Note that these periodicities are different
from those of the fractional revivals,
and thus a new level of revivals commences
for $t > t_{\rm rev}$.
We also find that at the particular time
$t_{\rm frac} \approx \fr 1 6 t_{\rm sr}$,
a single wave packet forms that resembles
the initial wave packet more closely than the
full revival does at time $t_{\rm rev}$,
i.e., a superrevival occurs.

In Refs.\ \cite{srsr2},
we have given theoretical proofs for the
periodicity and occurrence times of the
superrevivals.
We have looked at examples for large values
of $\bar n$ that illustrate the structure
of the fractional and full superrevivals clearly.
This structure may be seen as well in examples
with smaller values of $\bar n$.
Figure 1 shows the square of the autocorrelation
function for hydrogen with ${\bar n} = 48$ and $\si = 1.5$.
In this case,
the full revival is at $t \approx t_{\rm rev} \simeq 0.538$ nsec.
For times greater than this,
one can observe a fractional superrevival at $t \approx \frac 1 {12}
t_{\rm sr} \simeq 1.61$ nsec
with autocorrelation periodicity
$T_{\rm frac} \approx \frac 1 4 t_{\rm rev}$
and a full superrevival at
$t \approx \frac 1 6 t_{\rm sr} \simeq 3.23$ nsec
with autocorrelation periodicity
$T_{\rm frac} \approx \frac 1 2 t_{\rm rev}$.
The size of the peak in the autocorrelation function shows
that the superrevival resembles the initial wave packet
more closely than does
the revival wave packet at $t \approx t_{\rm rev}$.

\begin{figure}[t]
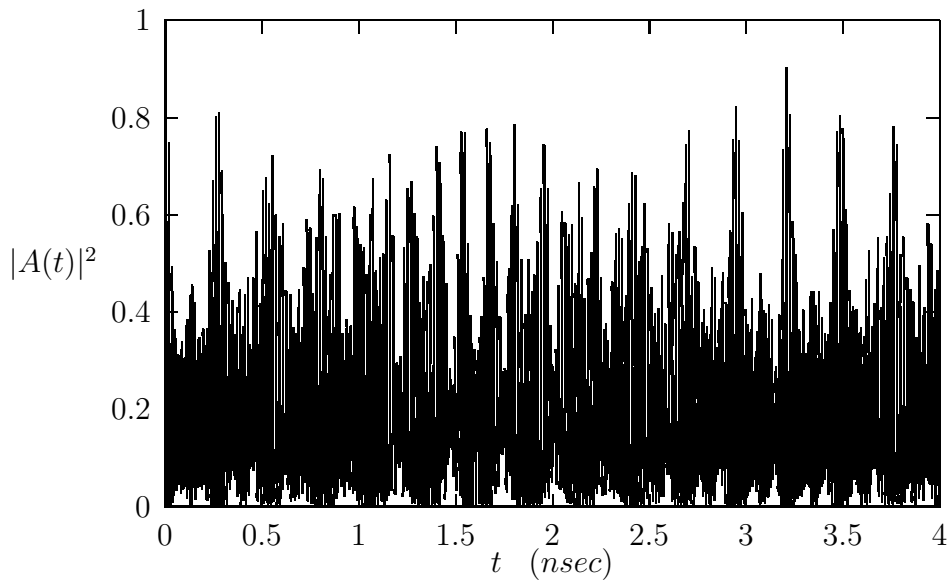

%\input{fig.tex}
% GNUPLOT: LaTeX picture
\setlength{\unitlength}{0.240900pt}
\ifx\plotpoint\undefined\newsavebox{\plotpoint}\fi
\sbox{\plotpoint}{\rule[-0.200pt]{0.400pt}{0.400pt}}%

\bigskip\bigskip
\caption{The absolute square of the autocorrelation function for a
Rydberg wave packet with ${\bar n} = 48$ and $\si = 1.5$
is plotted as a function of time in nanoseconds.}
\end{figure}

It appears likely that an experiment can be performed to detect
the long-time effects described in this talk.
One possibility is
to use the pump-probe method of detection
for Rydberg wave packets with
${\bar n} \approx 45$ -- $50$.
This is experimentally feasible,
provided a delay line of 3 -- 4 nsec
is installed in the apparatus.
This should permit detection of both
full and fractional superrevivals.

\vglue 0.4cm
{\bf\noindent 3. Radial Squeezed States}
\vglue 0.2cm

In this part of the talk,
we consider a description of radial Rydberg wave packets
as a type of squeezed state.
The theoretical analyses performed to date
on radial wave packets rely on
established tools such as expansion in eigenstates,
numerical methods, perturbation theory, and/or
the WKB approximation.
However,
the initial localization of the packet suggests
it might be profitably described in terms of
some kind of coherent state.
Standard approaches along these lines
either run into substantial technical difficulties
or generate objects that do not match the behavior
of p-state Rydberg atoms excited by short-pulsed lasers
with no external fields.

To construct the squeezed states
appropriate for a description of radial Rydberg wave packets,
we have adopted a procedure used in
Refs.\ \cite{ni} in the context of the construction
of `minimum-uncertainty coherent states.'
The idea is to change variables from $r$ and $p_r$
to a new set,
$R$ and $P$,
chosen to have oscillatory dependence on a suitable variable.
The similarities between the ensuing equations and
the usual quantum harmonic oscillator are sufficient
to render possible an analytical construction of
our candidate Rydberg wave packets.
Our method generates a three-parameter family
of radial squeezed states
\beq
\psi (r) =
\fr {(2 \ga_0 )^{2 \al + 3}} {\Ga (2 \al + 3)}
r^{\al} e^{-\ga_0 r} e^{-i \ga_1 r}
\quad .
\label{ras}
\eeq

For purposes of comparison of our radial squeezed states
with other theory and experiment,
we determine the parameters
$\al$, $\ga_0$, and $\ga_1$ by fixing the form of the
packet at the first pass through the classical apsidal point
by the conditions
$\vev{p_r} = 0$,
$\vev{r} = r_{\rm out}$,
$\vev{H} = E_{\bar n}$,
where
$r_{\rm out}$ is
the outer apsidal point of the orbit
and
$E_{\bar n} = - 1/2 \bar n^{2}$
is the energy of the dominant state
among those excited by the short laser pulse.

We have shown that these radial squeezed states may
be used as an initial wave function to model the motion of a
wave packet produced by a short laser pulse.
The time evolution of the radial squeezed states
exhibits the expected revival structure as well
as the oscillations in the uncertainty product
that are characteristic of a squeezed state.

\vglue 0.4cm
{\bf\noindent 4. Wave Packets in Alkali-Metal Atoms}
\vglue 0.2cm

All of the results described above for hydrogen may be
rederived in the context of
supersymmetry-based quantum-defect theory (SQDT)
\cite{konb}.
Since SQDT wave functions form a complete and
orthonormal set with the correct eigenenergies for
an alkali-metal atom,
the expansion of the energy for
a Rydberg wave packet may be carried out in
this context with the energies
$E_{n^\ast} = - 1/{2 n^{\ast 2}}$,
where $n^\ast = n - \de (l)$
and $\de (l)$ is an asymptotic quantum defect
for an alkali-metal atom.
In this case,
the Taylor expansion is
carried out around a noninteger central
value $N^\ast$ that may or may not be on resonance.
For the off-resonance case,
the noninteger part of $N^\ast$ consists of
two parts:  one from the quantum defect and another
from the laser detuning.
We may therefore consider quite generally the
question of how the effects of quantum defects
differ from those of a laser detuning in the
evolution of the wave packet.

In Ref.\ \cite{compare},
we have shown
that the effects of the quantum defects
are different from those of the laser detuning.
This difference arises because a constant shift
in the laser detuning is equivalent to a constant
shift for all energy levels,
whereas a constant shift in the quantum defect
would correspond to varying shifts among the
energy levels.

\vglue 0.4cm
{\bf\noindent 5. References}
\vglue 0.2cm


\begin{thebibliography}{xx}
\bibitem{note}
The reader is referred to Refs. \cite{srsr2,compare}
for additional references to background material.
\end{thebibliography}
\end{document}